\documentclass[12pt,oneside,final,a4paper]{article}

\title{Contextual viewpoint to quantum stochastics}
\author{Andrei Khrennikov\\
International Center for Mathematical Modeling\\
 in Physics and Cognitive Sciences,\\
 MSI, University of V\"axj\"o, S-35195, Sweden}
\begin{document}
\maketitle

\begin{abstract}
We study the role of context, complex of physical conditions,
in  quantum as well as classical experiments. It is shown that 
by taking into account contextual dependence of experimental
probabilities we can derive the quantum rule for the addition
of probabilities of alternatives. Thus we obtain quantum interference
without applying to wave or Hilbert space approach. The Hilbert
space representation of contextual probabilities is obtained
as  a consequence of the elementary geometric fact:
$\cos$-theorem. By using another fact from elementary algebra
we obtain complex-amplitude representation of probabilities.
Finally, we found contextual origin of noncommutativity
of incompatible observables.

\end{abstract}



 
 
 
\section{Introduction}

It is well known that the classical rule for the addition of probabilistic alternatives:
\begin{equation}
\label{F1}
P=P_1+P_2
\end{equation}
does not work in experiments with elementary particles. Instead of this rule, we have to use quantum rule:

\begin{equation}
\label{F2}
P=P_1+P_2+2\sqrt{P_1P_2}\cos\theta.
\end{equation}

The classical rule for the addition of probabilistic alternatives is perturbed by so called 
interference term. The difference between `classical' and `quantum' rules was (and is) the source of
permanent discussions as well as various misunderstandings and mystifications, see e.g. [1]-[12]
for general
references. We just note that the appearance
of the interference term was the source of the wave-viewpoint to the theory of elementary particles.
At least the notion of superposition of quantum states was proposed as an attempt to explain the appearance
of a new probabilistic calculus in the two slit experiment, see, for example, Dirac's book [1] on historical 
analysis of the origin of quantum formalism. We also mention that Feynman interpreted (\ref{F2}) as the evidence
of the violation of the additivity postulate for `quantum probabilities', [5].

In particular, this induced the viewpoint that there are some special 
`quantum' probabilities that differ essentially from ordinary `classical' probabilities. We also remark that 
the orthodox Copenhagen interpretation of quantum formalism is just an attempt to explain (\ref{F2}) without 
to apply to mysterious `quantum probabilities'. To escape the use of a new probabilistic calculus, we could 
suppose that, e.g. electron participating in the two slit experiment is in the superposition of passing through
both slits.

The role of an experimental {\it context} - a complex of physical conditions - in a
quantum measurement was discussed in the details already at the first stage of development 
of quantum theory. N. Bohr [3] pointed out that the experimental arrangement plays the crucial role
in quantum theory. W. Heisenberg also paid the large attention to the role of an 
experimental context (including the derivation of the uncertainty principle), see [2].
For Heisenberg, an experimental context was merely a source of perturbations - different
contexts produced different kinds of perturbations.  Dirac's approach to contextualism was 
similar to Heisenberg's viewpoint, see [1].  It is important
to note that Bohr and Heisenberg-Dirac contextual views  differed essentially. Heisenberg and Dirac
had classical-like viewpoint to the role of context as a source of perturbations induced by
force-like interactions. This was a kind of Newtonian scenarios. Bohr presented essentially more
general viewpoint to context. Bohr's contextualism was not based on the interpretation of context
as a source of force-like perturbations. In principle, there may be no force-like interaction
at all. Bohr's context is a kind of geometry, {\it experimental geometry.} The only difference
between ordinary geometry and contextual geometry  is that the first one is an individual,
point , model of physical reality and the second one is a statistical model of physical reality.

At the first stage of the development of quantum formalism contextualism was discussed
merely in the general philosophic framework. Of course, such a discussion played the great
role in creation of foundations of quantum mechanics. In particular, Bohr's {\it complementarity principle}
was created on the basis of contextual discussions. Unfortunately, contextualism
did not provided a mathematical model that could explain exotic features of quantum experimental
statistics. In particular, there was not found contextual purely classical probabilistic
derivation of quantum probabilistic rule (\ref{F2}). This inability to obtain interference for
quantum particles considered as classical-realistic objects - but moving in varying contextual
geometries - induced a "romantic viewpoint" to quantum systems: wave features, absence of trajectories,...
Moreover, it was one of the roots of so called Orthodox Copenhagen Interpretation of quantum
mechanics (not mix with Bohr's interpretation!): {\it a wave function provides
the complete description of an individual quantum system.} By this interpretation
a wave function was associated not with the complex of physical conditions, context,
(as it should be according to N. Bohr), but with an individual physical quantum system.

There are many reasons why contextualism was not successive in deriving of 
quantum statistics. One of problems was of purely mathematical character. The standard 
probabilistic formalism based on {\it Kolmogorov's axiomatics} [13], 1933, was a fixed context
formalism. This conventional probabilistic formalism  does not provide rules of operating 
with probabilities calculated for different contexts. However, in quantum theory we have to
operate with statistical data obtained for different complexes of physical conditions,
contexts. In fact, this context dependence of probabilities as the origin of the superposition
principle was already discussed by W. Heisenberg [2]; unfortunately, only in quite general
and rather philosophic framework. Later contextual dependence of quantum probabilities
was intensively investigated from various viewpoints [14]-[19]. We also mention that the powerful contextual
machinery was developed in the operational framework, see [8],[9];
in particular, by using the theory based on POV-measures, see [20], [21]. However, a simple contextual
derivation of quantum probabilistic rule (\ref{F2}) was not found. Typically contextual papers
explained that contextual transitions could violate classical probabilistic
rule (\ref{F1}), see e.g. L. Ballentine for clear and simple presentation in [19].
However, it was not clear why we have interference rule (\ref{F2}) in quantum formalism
and why we use the Hilbert space calculus of quantum probabilities.

Such a classical probabilistic derivation of (\ref{F2}) was given in author's paper [22],
see also [23]. In these papers we used Heisenberg-Dirac contextualism: experimental contexts
produces perturbations of physical variables. By using magnitudes of statistical perturbations
we derived (\ref{F2}). We should mention two disadvantages of this derivation:
1) the use of force-like perturbation picture of physical reality; 2) the use of von Mises'
frequency probability model [24]. Restrictions of old-fashioneds perturbation approach
becomes more evident in the process of the development of quantum theory. R. von Mises frequency
approach induces quite complicated manipulations with frequencies [22], [23] that are, in fact, irrelevant
to our physical considerations. Therefore we tried to present Bohr-like contextual derivation
of (\ref{F2}) and without to go to the frequency level. We did this in the preprint [25].

In this paper we present a modified variant of derivation [25] of (\ref{F2}). Then we 
present contextual derivation of the superposition principle as superposition of waves 
of contexts. Starting with (now classically derived) (\ref{F2}) we get vector space (and complex amplitude)
representation of contextual probabilistic calculus.

\section{The role of a complex of physical conditions}

In fact, probabilities in (\ref{F2}) in quantum experiments 
are determined by at least three different contexts,
${\cal S}, {\cal S}_1, {\cal S}_2.$ We illustrate this situation by following fundamental example.

{\bf Example.} (Two slit experiment) In the two slit experiment rule (\ref{F2}) is induced by
combining of statistical data obtained in three different experiments: 
both slits are open; only $j$th slit is open, $j=1,2.$ The main distinguishing feature
of statistical data obtained in these three experiments in the following one. 
By combining by (\ref{F1}) data
obtained in experiments in that only one of slits is open we do not get the probability distribution for
data obtained in the experiment in that both slits are open. On the other hand we never observe a particle
that passes through both slits simultaneously - it would be observed passing the first or second 
slit. There  is no the direct observation of particle splitting. As each particle passes  only one of slits,
we have the standard case of alternatives. Thus we should use conventional rule (\ref{F1}) for
the addition of probabilities of alternatives. This disagreement between experimental statistical
data and the rule of conventional probability theory looks as a kind of paradox. The traditional
solution of this paradox is the use of the wave model for elementary particles.

We now perform detailed contextual analysis for the two slits experiment. We consider the following
complexes of physical conditions, contexts.

${\cal S}=${both slits are open}, ${\cal S}_j=${only jth slit is open}, $j=1,2.$ 

In fact, probabilities in (\ref{F2}) are related to these three contexts.
Thus $P=P_{{\cal S}}(A)$ and $P_j= P_{{\cal S}_j/{\cal S}} P_{{\cal S}_j}(A), j=1,2.$ 

Here we use various context-indexes. The $P_{{\cal S}}(A), P_{{\cal S}_j}(A)$ denote probabilities
of an event $A$ with respect to various contexts. The coefficients 
$P_{{\cal S}_j/{\cal S}}, j=1,2,$ have another
meaning. In general these
are not probabilities of ${\cal S}_j$ with respect to the context ${\cal S}$
(besides some very special, "classical", situations), because the context
${\cal S}_j$ in general is not an event for the context ${\cal S}.$
The $P_{{\cal S}_j/{\cal S}}, j=1,2,$
are kinds of {\it balance probabilities}. These are proportion coefficients in splitting
of the ensemble $S$ prepared on the basis of the complex of physical conditions ${\cal S}$
into ensembles $S_j$ prepared on the basis of the complexes of physical conditions ${\cal S}_j:$
\begin{equation}
\label{BP}
P_{{\cal S}_j/{\cal S}}= \frac{N_j}{N},
\end{equation}
where $N$ is the number of elements in $S$ and $N_j$ 
is the number of elements in $S_j,$ see [22], [23] for the details.
We remark (and it is important for our
further considerations) that we have the following balance condition:
\begin{equation}
\label{BC}
P_{{\cal S}_1/{\cal S}} + P_{{\cal S}_2/{\cal S}} =1.
\end{equation}
The balance condition has the following meaning: the total number of particles that
arrives to the registration screen when both slits are open equals (in the average)
to the sum of corresponding numbers when only one of the slits is open. So by closing e.g. the first
slit we do not change the number of particles that pass the second slit (in the average).
In fact, (\ref{BC}) gives the right description of alternative-situation in the two slit
experiment. It is not related to alternative passing of slits by a particle in the experiment
when both slits are open. This equation describes alternative sharing of particles between
two preparation procedures: $j$th slit is open, $j=1,2.$

However, the balance probabilities $P_{{\cal S}_j/{\cal S}}$  would not play 
so important role in our considerations. The crucial role will be played by contextual
probabilities $P_{{\cal S}}(A), P_{{\cal S}_j}(A).$

The conventional probability theory says, see e.g. [26] 
that we have:
\begin{equation}
\label{FT}
P(A)= P({\cal S}_1) P(A/{\cal S}_1) + P({\cal S}_2) P(A/{\cal S}_2).
\end{equation}
if ${\cal S}_1\cap {\cal S}_2 = \emptyset$ and ${\cal S}_1\cup {\cal S}_2 ={\cal S}.$
This is the well known {\it formula of total probability.}
In many considerations (including works of fathers 
of quantum mechanics, see e.g. P. Dirac [1], see also R. Feynman [5]) people 
set $P=P(A)$ and $P_j= P({\cal S}_1) P(A/{\cal S}_1).$
Finally, they get the contradiction between conventional
probabilistic rule (\ref{F1}) and statistical data obtained
in the interference experiments and described by quantum rule (\ref{F2}).

We would like to discuss physical and mathematical
assumptions used for the derivation of (\ref{FT}).
The main physical assumption is that this formula is  derived for one fixed context ${\cal S}$
(in the mathematical formalism - for one fixed Kolmogorov probability
space). To be precise, we have to write this formula as 
\begin{equation}
\label{FT1}
P_{{\cal S}}(A)= P_{{\cal S}}({\cal S}_1)  P_{{\cal S}}(A/{\cal S}_1)
+ P_{{\cal S}}({\cal S}_2)  P_{{\cal S}}(A/{\cal S}_2).
\end{equation}
It is also important that contexts ${\cal S}_1,  {\cal S}_2$ 
can be realized as elements of the field of events corresponding
to the context ${\cal S}.$
Thus we would get the contradiction between classical rule (\ref{F1}) and
quantum rule (\ref{F2}) only if assume that balance probabilities
$P_{{\cal S}_j/{\cal S}}$ can be interpreted as 
$P_{{\cal S}}$-probabilities, $P_{{\cal S}_j/{\cal S}}=P_{{\cal S}} ({\cal S}_j)$ 
and contextual probabilities $P_{{\cal S}_j}(A)$
as conditional probabilities with respect to the context ${\cal S}, P_{{\cal S}_j}(A)=
P_{{\cal S}}(A/{\cal S}_j).$  Here the contextual 
probabilities are given by Bayes' formula
(the additional postulate of Kolmogorov's probability theory):

$P_{{\cal S}_j}(A)= P_{{\cal S}}(A \cap {\cal S}_j)/ P_{{\cal S}}({\cal S}_j) .$

In general, there are no reasons to assume that new complexes of conditions ${\cal S}_j$
are "so nice" that new probability distributions are given by the Bayes' formula.
Thus, in general, the formula of total probability can be violated
when we combine statistical data obtained for a few distinct contexts.
In particular,
this formula does not hold true in statistical experiments with elementary particles.
The right hand side of it is perturbed by the interference term. This  explanation
is well known in contextual community.  We could only be surprised why R. Feynman 
was so surprised by such "exotic behaviour" of quantum probabilities in the two
slit experiment, see [5]. 

{\bf Remark.} {\small We can discuss the same experiment in the standard framework of 
preparation/measurement procedures. The context ${\cal S}$ produced an ensemble of 
particles $S$ that passed through the screen with slits  when both slits are open. A measurement 
procedure is the measurement of the position of a particle on the registration screen.
In the same way the context ${\cal S}_j$ produced an ensemble of 
particles $S_j$ that passed through the screen with slits when only $j$th slit is open.}

{\bf Remark.}  {\small  I recently discovered that, in fact, there is no contradiction 
between Kolmogorov's approach [13] and the contextual approach.
\footnote{I would like to thank Prof. A. Shiryaev (the former student of A. N. Kolmogorov) who explained this to me.}
It is the mysterious fact that a few generations of readers (including the author of this paper) 
did not pay attention to section 2 of Kolmogorov's book [13]. There Kolmogorov said directly 
that all probabilities of events are related to concrete complexes of conditions. Moreover, in his 
paper [27] he even used the symbol $P(A|{\cal S}),$ where ${\cal S}$ is a complex of conditions,
- an analogue of our symbol $P_{{\cal S}}(A).$}

\section{Interference term as the measure of statistical deviations due to the context transition.}
The following simple considerations gives us the derivation of 
quantum probabilistic transformation (\ref{F2}) in the 
classical probabilistic framework.  

Let ${\cal S}$ and ${\cal S}_j, j=1,2,$ be three
different complexes of conditions. We consider the transformation of probabilities induced by transitions 
from one complex of conditions to others: 
\begin{equation}
\label{CT}
{\cal S} \rightarrow {\cal S}_1\; \mbox{and}\; {\cal S} \rightarrow {\cal S}_2.
\end{equation}

We start with introducing of balance probabilities,
$P_{{\cal S}_j/{\cal S}}.$  These are proportional coefficients for numbers of physical systems
obtained after preparations under the complexes of physical conditions ${\cal S}$ and ${\cal S}_j.$
If (starting with the same number of particles)  we get $N$ and $N_j$ systems after  ${\cal S}$
and ${\cal S}_j$ preparations, respectively, then $P_{{\cal S}_j/{\cal S}}$  are defined by
(\ref{BP}). We assume that balance probabilities satisfy to balance equation (\ref{BC}).
This is quite natural conditions: splitting (\ref{CT}) of the context ${\cal S}$ 
induces just  sharing of physical systems produced by a source. We have already
discussed this balance in the two slit experiment. The same situation we have 
in neutron interferometry  for the balance between the numbers of particles coming to detectors
when both paths are open and when just one of the paths is open.

We introduce  the {\it measure of statistical perturbations}
$\delta$ induced by context transitions:

$\delta(A;{\cal S}; {\cal S}_j)=P_{{\cal S}_1/{\cal S}} [P_{{\cal S}}(A)-
P_{{\cal S}_1}(A)]+ P_{{\cal S}_2/{\cal S}} [P_{{\cal S}}(A)-
P_{{\cal S}_2}(A)].$

This quantity describes the deformation of probability distribution $P_{{\cal S}}$
due to context transitions.
\medskip

By using balance equation (\ref{BC}) we get:

$P_{{\cal S}}(A)= P_{{\cal S}}(A) P_{{\cal S}_1/{\cal S}}  + P_{{\cal S}}(A)P_{{\cal S}_2/{\cal S}} .$

Thus we get:
\begin{equation}\label{F3}
P_{{\cal S}}(A)= P_{{\cal S}_1/{\cal S}}  P_{{\cal S}_1}(A)+ P_{{\cal S}_2/{\cal S}}  P_{{\cal S}_2}(A)+
\delta(A;{\cal S}, {\cal S}_j).
\end{equation}

Transformation (\ref{F3}) is the most general form of probabilistic transformations due to context transitions. 

There is {\it{the correspondence principle}} between context unstable and (`classical') context
stable transformations: If ${\cal S}_j \rightarrow {\cal S}, j=1,2,$ 
i.e., $\delta(A; {\cal S}, {\cal S}_j) \rightarrow 0$, 
then contextual probabilistic transformation (\ref{F3}) coincides (in the limit) with 
the conventional formula of total probability.

The perturbation term $\delta(A; {\cal S}, {\cal S}_j)$ depends on absolute magnitudes of probabilities. 
It would be natural to introduce normalized coefficient of the context transition 

$\lambda(A; {\cal S}, {\cal S}_j)=
\frac{\delta(A; {\cal S}, {\cal S}_j)}{2\sqrt{P_{{\cal S}_1/{\cal S}}  P_{{\cal S}_1}(A)P_{{\cal S}_2/{\cal S}}  P_{{\cal S}_2}(A)}},$

that gives the relative measure of statistical deviations due to the transition from one complex of conditions, 
${\cal S},$ to others, ${\cal S}_j.$ Transformation (\ref{F3}) can be written in the form:

\begin{equation}
\label{F4}
P_{{\cal S}}(A)= \sum_{j=1,2} P_{{\cal S}_j/{\cal S}}  P_{{\cal S}_j}(A)+
2 \sqrt{P_{{\cal S}_1/{\cal S}}  P_{{\cal S}_1}(A)P_{{\cal S}_2/{\cal S}}  P_{{\cal S}_2}(A)} \lambda(A; {\cal S}, {\cal S}_j)\;.
\end{equation}

In fact, there are two possibilities:

1). $|\lambda(A; {\cal S}, {\cal S}_j)|\leq 1;$

2). $|\lambda(A; {\cal S}, {\cal S}_j)|\geq 1.$

In both cases it is convenient to introduce a new context transition parameter 
$\theta= \theta(A; {\cal S}, {\cal S}_j)$ and represent
the context transition coefficient in the form:
\[\lambda(A; {\cal S}, {\cal S}_j)=\cos \theta (A; {\cal S}, {\cal S}_j) , \theta \in [0, \pi];\] 
and
\[\lambda(A; {\cal S}, {\cal S}_j)=\pm\cosh \theta (A; {\cal S}, {\cal S}_j), \theta \in [0, \infty),\]
respectively.

We have two types of probabilistic transformations induced by the transition 
from one complex of conditions to another: 

\begin{equation}
\label{F5}
P_{{\cal S}}(A)= \sum_{j=1,2} P_{{\cal S}_j/{\cal S}}  P_{{\cal S}_j}(A) +
2 \sqrt{P_{{\cal S}_1/{\cal S}}  P_{{\cal S}_1}(A)P_{{\cal S}_2/{\cal S}}  P_{{\cal S}_2}(A)} \cos \theta(A; {\cal S}, {\cal S}_j)\;.
\end{equation}
\begin{equation}
\label{F6}
P_{{\cal S}}(A)= \sum_{j=1,2} P_{{\cal S}_j/{\cal S}}  P_{{\cal S}_j}(A) \pm
2 \sqrt{P_{{\cal S}_1/{\cal S}}  P_{{\cal S}_1}(A)P_{{\cal S}_2/{\cal S}}  P_{{\cal S}_2}(A)} 
\cosh \theta(A; {\cal S}, {\cal S}_j)\;.
\end{equation}

We derived quantum probabilistic rule (\ref{F2})  in the classical probabilistic framework 
(in particular, without any reference to superposition of states) by taking into account context
dependence of probabilities.

Relatively large statistical deviations are described by transformation (\ref{F6}). Such transformations do not 
appear in the conventional formalism of quantum mechanics. In principle, they could be described by so called 
{\it{hyperbolic quantum mechanics}}, [28].

{\bf Conclusion.} For each fixed context (experimental arrangement), we have CLASSICAL STATISTICS.
CONTEXT TRANSITION induces interference perturbation
of the conventional rule for the addition of probabilistic alternatives.

\section{Linear algebra for probabilities, complex amplitudes}

One of the main distinguishing features of quantum theory
is the Hilbert space calculus for probabilistic amplitudes.
As we have already discussed, this calculus is typically associated
with wavelike (superposition) features of quantum particles.
We shall show that, in fact, the Hilbert space representation of
probabilities was merely a mathematical discovery. Of course, this discovery
simplifies essentially calculations. However, this is pure mathematics;
physics is related merely to the derivation of
quantum interference rule (\ref{F2}). 

The crucial point was 
the derivation (at the beginning purely experimental) of transformation (\ref{F2})
connecting probabilities with respect to three different contexts. In fact, linear algebra
can be easily derived from this transformation. Everybody familiar with the elementary geometry
will see that (\ref{F2}) just the well known $\cos$-theorem. This is the rule to find
the third side in a triangle  if we know
lengths of two other sides and the angle $\theta$ between them:
$$
c^2 =a^2 + b^2 - 2 ab \cos \theta\;.
$$
or if we want to have "+" before $\cos$ we use so called {\it parallelogram law:}
\begin{equation}
\label{P}
c^2 = a^2 + b^2 + 2 ab \cos \theta\;.
\end{equation}
Here $c$ is the diagonal of the parallelogram with sides $a$ and $b$ and the angle $\theta$
between these sides. Of course, the parallelogram law is  just the law of linear
(two dimensional Hilbert space) algebra: for finding the length $c$ of the sum ${\bf c}$
of vectors ${\bf a}$ and ${\bf b}$ having lengths $a$ and $b$ and the angle $\theta$ between them.

We also can introduce 
complex waves by using the following 
elementary formula: 
\begin{equation}
\label{TTT}
a^2 + b^2 + 2 ab \cos\theta=|a+ b e^{i\theta}|^2\; .
\end{equation}
Thus the context transitions 
${\cal S}\rightarrow {\cal S}_j$ can be described by the wave:
$$
\varphi=\sqrt{P_{{\cal S}_1/{\cal S}}  P_{{\cal S}_1}(A)} 
+ \sqrt{P_{{\cal S}_2/{\cal S}}  P_{{\cal S}_2}(A)}e^{i\theta(A; {\cal S}, {\cal S}_j)}.
$$

\section{`Classical' probabilistic derivation of the superposition
principle for wave functions in the two slit experiment}

We shall study in more details the possibility of
contextual (purely classical) derivation of the superposition
principle for complex probability amplitudes, `waves', 
in the two slit experiment. We consider one dimensional
model. It could be obtained by considering the distribution of 
particles on one fixed straight line, very thin strip. It is supposed
that the source of particles is symmetric with respect to slits and 
the straight line (on the registration screen) pass through the center
of the screen. This geometry implies that 
$P_{{\cal S}_j/{\cal S}} =1/2, j=1,2.$ By the symbol $A_x, x \in {\bf R},$
is denoted the event of  the registration of a particle at the point $x$ 
ofd the straight line. We set:

$ p(x)= P_{{\cal S}}(A_x)$ and $p_j(x)= P_{{\cal S}_j}(A_x), j=1,2,$

where contexts ${\cal S}$ and ${\cal S}_j$ were defined in Example 1.
By using
(\ref{F5}) we get:

$p(x)= \frac{1}{2} [p_1(x)+p_2(x)+ 2 \sqrt{p_1(x)p_2(x)} \cos \theta(x)] .$

By using (\ref{TTT}) we represent this probability as the square of a complex
amplitude, $p(x)= \vert \phi(x)\vert^2,$

where 
\begin{equation}
\label{S}
\phi(x)= \frac{1}{\sqrt{2}} (e^{i \theta_1(x)}\sqrt{p_1(x)} + 
e^{i \theta_2(x)}\sqrt{p_2(x)})
\end{equation}
and phases $\theta_j(x)$ are chosen in such a way that the phase shift
$\theta_1(x) - \theta_2(x) = \theta (x).$ We also introduce  complex amplitudes
for probabilities $p_j(x): \; \phi_j(x)= \frac{1}{\sqrt{2}} e^{i \theta_j(x)}\sqrt{p_j(x)}.$
Here $p_j(x)= \vert \phi_j(x)\vert^2.$ The complex amplitudes are 
said to be {\it wave functions:}  $\phi(x)$ is the wave function on (the straight
line of) the registration screen for both slits are open;
$\phi_j(x)$ is the wave function on (the straight
line of) the registration screen for $j$th slit  is open.

Let us set $\xi(x)=\frac{\theta(x)}{h},$ where $h>0$ is some scaling factor.
We have:
$$
\phi(x)= \frac{1}{\sqrt{2}} (e^{\frac{i \xi_1(x)}{h}}\sqrt{p_1(x)} + 
e^{\frac{i \xi_2(x)}{h}}\sqrt{p_2(x)})\; \mbox{and}\; 
\phi_j(x)= \frac{1}{\sqrt{2}} e^{\frac{i \xi_j(x)}{h}}\sqrt{p_j(x)}.
$$
By choosing $h$ as the Planck constant we get a quantum-like representation
of probabilities. We recall that we did not use any kind of wave arguments.
Superposition rule (\ref{S}) was obtained in purely classical probabilistic
(but contextual!) framework.

Suppose now that $\xi$ depends linearly on $x: \xi_j(x)= \frac{{\bf p}_j x}{h}, \xi(x)= \frac{{\bf p} x}{h}, 
{\bf p} ={\bf p}_1 - {\bf p}_2.$ Under such an assumption we shall get interference
of two `free-waves' corresponding to momentums ${\bf p}_1$ and ${\bf p}_2.$ Of course, this
linearity could not be extracted from  our general probabilistic considerations. This is a
consequence of the concrete geometry of the experiment.

\section{The coefficient of context transition as the measure of incompatibility of physical
observables}

We now consider the relation between the coefficient of context transition
(the measure of statistical deviations due to the change of complex of physical conditions)
and incompatibility of physical observables in quantum mechanics (noncommutativity of
corresponding operators). As everywhere in this paper, we consider dichotomic observables.
Each event $A$ generates the dichotomic variable $a:$ $a=a_1$ if $A$ occurs and $a=a_2$ if $A$ does not
occur. Values $a_1$ and $a_2$ do not play any role in our considerations; in principle,
we can consider the case $a_1=0$ and $a_2=1.$

{\bf Definition.} {\it A physical observable $a$ is incompatible with a pair ${\cal S}_1, {\cal S}_2$ of contexts 
if there exists  a context ${\cal S}$  such
that $\delta(a=a_i; {\cal S}; {\cal S}_j)\not=0.$}

In such a case a transition from the complex of physical conditions ${\cal S}$ to complexes
${\cal S}_j$ induces non-negligible statistical deviations for $a$-measurements. It is not the same
to measure $a$ under the complex of conditions ${\cal S}$ or ${\cal S}_j.$\footnote{
We need to consider two complexes ${\cal S}_1$ and ${\cal S}_2,$ because we would like to 
consider another dichotomic variable $b$ connected to these contexts.}

We shall demonstrate that the incompatibility of physical observables in quantum mechanics
is just a particular case of contextual incompatibility. 

Let ${\cal H}$ be the two dimensional Hilbert space. Rays of this space represent 
some class of complexes of physical conditions. 
Let the dichotomic variable $a$ be represented  by a self-adjoint operator (symmetric
matrix) $\hat{a}.$ We remark that we can associate with any physical observable $a$ 
two complexes of conditions ${\cal S}_1^a, {\cal S}_2^a$ 
namely contexts corresponding to eigenvectors $\phi_1^a, \phi_2^a$ of  $\hat{a}.$ 
The ${\cal S}_j^a$ describes the filter with respect to the value $a=a_j$.

Let us consider other dichotomic physical observable $b=b_1, b_2.$ It is represented
by a self-adjoint operator $\hat{b}$ with eigenvectors $\phi_1^b, \phi_2^b.$ 
These eigenvectors represent contexts ${\cal S}_1^b, {\cal S}_2^b$ 
(filtrations corresponding to $b=b_1$ and $b=b_2,$ respectively).

{\bf Theorem.} {\it Quantum physical observables $a$ and $b$ are incompatible
(i.e., corresponding operators do not commute) iff the observable $a$ is incompatible
with the contexts ${\cal S}_j^b$ or vice versa:}

$\delta(a=a_i; {\cal S}; {\cal S}_j^b)\not=0$ or $\delta(b=b_i; {\cal S}; {\cal S}_j^a)\not=0.$

{\bf Proof.} Let ${\cal S}$ be an arbitrary quantum context. Thus it can be represented
by a normalized vector $\phi \in {\cal H}.$ We have:
$$
\delta(a=a_i; {\cal S}; {\cal S}_j^b)=
$$
$$
P_{{\cal S}_1/{\cal S}} [P_{{\cal S}}(a=a_i)-
P_{{\cal S}_1}(a=a_i)]+ P_{{\cal S}_2/{\cal S}} [P_{{\cal S}}(a=a_i)-
P_{{\cal S}_2}(a=a_i)]
 $$
 $$
 = \vert (\phi, \phi_1^b)\vert^2( \vert (\phi, \phi_i^a)\vert^2 - \vert (\phi_i^a, \phi_1^b)\vert^2)
 + \vert (\phi, \phi_2^b)\vert^2( \vert (\phi, \phi_i^a)\vert^2 - \vert (\phi_i^a, \phi_2^b)\vert^2).
 $$
 We have $(\phi, \phi_j^b)= k_j e^{i \xi_j}, (\phi_j^b, \phi_i^a) = k_{ji} e^{i \xi_{ji}},$
 where $k_j, k_{ij} \geq 0.$ We get:
$$
\delta(a=a_i; {\cal S}; {\cal S}_j^b)= 2 k_1 k_2 k_{1i} k_{2i} \cos \theta_i,
$$
where $\theta_i= \xi_2-\xi_1 + \xi_{2i}- \xi_{1i}.$

a). Let $[\hat{a}, \hat{b}]=0.$ Then $k_{12}= k_{21}=0.$  Hence
$\delta(a=a_i; {\cal S}; {\cal S}_j^b)=0.$

b). Let $[\hat{a}, \hat{b}] \not =0.$ Then $k_{12}, k_{21}\not =0.$
Let $k_1,k_2 >0$ be arbitrary constants such that 
$k_1^2+k_2^2=1.$ We choose a context ${\cal S}$ that is described
by the state:

$\phi= \sqrt{k_1} e ^{i\xi_{21}} \phi_1^b +\sqrt{k_2} e ^{i\xi_{11}} \phi_2^b.$

Here $\theta=0$ and, hence,

$\delta(a=a_i; {\cal S}; {\cal S}_j^b)=2 k_1 k_2 k_{i1} k_{i2} >0.$

I would like to thank  L. Ballentine,  W. De Myunck, 
D. Mermin, A. Peres  for fruitful (rather critical) discussions. 

{\bf References}

[1] P. A. M.  Dirac, {\it The Principles of Quantum Mechanics}
(Oxford Univ. Press, 1930).

[2] W. Heisenberg, {\it Physical principles of quantum theory.}
(Chicago Univ. Press, 1930).

[3]  N. Bohr, {\it Phys. Rev.,} {\bf 48}, 696-702 (1935).

[4] J. von Neumann, {\it Mathematical foundations
of quantum mechanics} (Princeton Univ. Press, Princeton, N.J., 1955).

[5] R. Feynman and A. Hibbs, {\it Quantum Mechanics and Path Integrals}
(McGraw-Hill, New-York, 1965).

[6] J. M. Jauch, {\it Foundations of Quantum Mechanics} (Addison-Wesley,
Reading, Mass., 1968).

[7] A. Peres, {\em Quantum Theory: Concepts and Methods} (Kluwer Academic
Publishers, 1994).

[8]  G. Ludwig, {\it Foundations of quantum mechanics,} v.1. Springer-Verlag, Berlin
(1983).

[9] P. Busch, M. Grabowski, P. Lahti, {\it Operational Quantum Physics}
(Springer Verlag, 1995).

[10] E. Beltrametti  and G. Cassinelli, {\it The logic of Quantum mechanics.}
(Addison-Wesley, Reading, Mass., 1981).

[11]  L. E. Ballentine, {\it Quantum mechanics} (Englewood Cliffs, 
New Jersey, 1989).

[12] A.Yu. Khrennikov, {\it Interpretations of 
probability} (VSP Int. Publ., Utrecht, 1999).

[13] A. N. Kolmogoroff, {\it Grundbegriffe der Wahrscheinlichkeitsrechnung.}
Springer Verlag, Berlin (1933); reprinted:
{\it Foundations of the Probability Theory}. 
Chelsea Publ. Comp., New York (1956).

[14] S. P. Gudder,  J. Math Phys., {\bf 25}, 2397 (1984); S. P. Gudder, N.
Zanghi, Nuovo Cimento B {\bf 79}, 291(1984).

[15] L. Accardi, Phys. Rep., {\bf 77}, 169(1981); 
L. Accardi, {\it Urne e Camaleoni: Dialogo sulla realta,
le leggi del caso e la teoria quantistica.} Il Saggiatore, Rome (1997).

[16] I. Pitowsky,  Phys. Rev. Lett, {\bf 48}, N.10, 1299(1982).
 
[17]  A. Fine,  Phys. Rev. Letters, {\bf 48}, 291 (1982);
P. Rastal, Found. Phys., {\bf 13}, 555 (1983).

[18] W. De Baere,  Lett. Nuovo Cimento, {\bf 39}, 234 (1984);
{\bf 25}, 2397(1984); 
 W. De Muynck, W. De Baere, H. Martens,
 Found. of Physics, {\bf 24}, 1589 (1994);

[19] L. Ballentine, Probability theory in quantum mechanics. {\it American
J. of Physics,} {\bf 54}, 883-888 (1986).

[20] E. B. Davies, {\it Quantum theory of open systems}, Academic Press, 
London (1976).

[21] A. S. Holevo, {\it Probabilistic and 
statistical aspects of quantum 
theory.} North-Holland, Amsterdam (1982).

[22] A. Yu. Khrennikov, {\it Ensemble fluctuations and the origin of
quantum probabilistic rule.} Rep. MSI, V\"axj\"o Univ., {\bf 90}, October (2000).

[23] A. Yu. Khrennikov, {\it Linear representations of probabilistic
transformations induced
by context transitions.} Preprint quant-ph/0105059, 13 May 2001;
to be published in {\it J. of Phys. A.}.

[24] R.  von Mises, {\it The mathematical theory of probability and
 statistics}. (Academic, London,  1964);

[25] A. Yu. Khrennikov, {\it `Quantum probabilities'  as context depending
probabilities.}
Preprint quant-ph/0106073, 13 June 2001.

[26] A. N. Shiryayev, {\it Probability.} (Springer, New York-Berlin-Heidelberg, 1991).

[27]   A. N. Kolmogorov, {\it Theory of Probability.} In series {\it Mathematics, its context, methods and role},
{\bf 2}, 252-290, Academy of Sc. of USSR, Steklov Math. Institute (in Russian).
 
[28] A. Yu. Khrennikov, {\it Hyperbolic Quantum Mechanics.} Preprint: quant-ph/0101002, 31 Dec 2000.

\end{document}